\documentclass{emulateapj}
\usepackage{apjfonts}
\slugcomment{Accepted for publication in the Astrophysical Journal}
\shorttitle{The Systematic Properties of X-ray Cavities}
\shortauthors{B\^{\i}rzan et al.}

\begin{document}

\title{A Systematic Study of Radio-Induced X-ray Cavities in Clusters, Groups,
and Galaxies}

\author{L. B\^{\i}rzan, D. A. Rafferty, B. R. McNamara}
\affil{Department of Physics and Astronomy, Ohio University, Clippinger Laboratories, Athens, OH 45701}
\author{M. W. Wise}
\affil{Center for Space Research, Building NE80-6015, Massachusetts Institute of Technology, Cambridge, MA 02139}
\and
\author{P. E. J. Nulsen}
\affil{Harvard-Smithsonian Center for Astrophysics, 60 Garden St., Cambridge,
MA 02138,}
\affil{School of Engineering Physics, University of Wollongong, Wollongong, NSW 2522, Australia }

\begin{abstract}
We present an analysis of sixteen galaxy clusters, one group and one galaxy
drawn from the \textit{Chandra} X-ray Data Archive. 
These systems possess prominent X-ray surface brightness depressions
associated with cavities or bubbles that were 
created by interactions between powerful radio sources and
the surrounding hot gas.  The central galaxies in these systems 
harbor radio sources with luminosities ranging
between $\sim 2\times 10^{38}-7\times 10^{44}~\mbox{ergs s$^{-1}$}$.
The cavities have an average
radius of $\sim 10~\mbox{kpc}$, and they lie at an average projected distance 
of $\sim 20~\mbox{kpc}$ from the
central galaxy.  The minimum energy associated with the
cavities ranges from $pV \sim 10^{55}~\mbox{ergs}$ 
in galaxies, groups, and poor clusters
to $pV \sim 10^{60}~\mbox{ergs}$ in rich clusters.  We evaluate the 
hypothesis that cooling in the hot gas can be quenched 
by energy injected into the surrounding gas by the rising bubbles.
We find that the instantaneous mechanical 
luminosities required to offset cooling 
range between $1 pV$ and $20 pV$ per cavity.
Nearly half of the systems in this study may have 
instantaneous mechanical luminosities large enough to balance cooling,
at least for a short period of time, if
the cavities are filled with a relativistic gas. 
We find a trend or upper envelope in the distribution 
of central X-ray luminosity versus 
instantaneous mechanical luminosity with the sense 
that the most powerful cavities are found in the most X-ray--luminous systems.
Such a trend would be expected if many of these  
systems produce bubbles at
a rate that scales in proportion to the cooling rate of the surrounding
gas.  Finally, we use
the X-ray cavities to measure the mechanical power of radio sources
over six decades of radio luminosity, 
independently of the radio properties themselves.  We find that the ratio of
the instantaneous mechanical (kinetic) luminosity 
to the 1.4~GHz synchrotron 
luminosity ranges from a few to roughly a thousand.    
This wide range implies that the 1.4 GHz synchrotron luminosity
is an unreliable gauge of the mechanical power of radio sources.
\end{abstract}


\section{Introduction}

The cooling time of the intracluster gas in the cores of many galaxy clusters
is shorter than 1~Gyr.  In the absence of heating, a  
``cooling flow'' \citep{fabi94} is established, in which the gas cools
below X-ray temperatures and accretes onto the central cluster
galaxy, where it accumulates in molecular clouds and forms stars.
{\it Chandra} images of cooling flow clusters have 
confirmed the existence of inwardly decreasing
temperature gradients and short central cooling times, which are
the distinguishing characteristics of a cooling flow.
However, moderate-resolution \textit{Chandra} and \textit{ASCA} spectra 
and high resolution \textit{XMM-Newton} spectra 
\citep[e.g.\ ][]{maki01,pete01,tamu01, kaast04} 
do not show the expected signatures of cooling below 2~keV,  reported 
to exist in lower
resolution data from the \textit{Einstein} and \textit{ROSAT} observatories.  
This discrepancy would be difficult to understand unless
the normal signatures of cooling below 2 keV are 
somehow suppressed, or if cooling is indeed occurring but at
rates that are generally factors of $5-10$ lower than 
expected \citep[e.g.\ ][]{mole01,boeh02,pete03}.
Several scenarios have
been suggested that may suppress the cooling flux
at low energies, and yet maintain the large cooling rates.  
The list includes differential absorption, efficient mixing, or
inhomogeneous metallicity distributions \citep[e.g.\ ][]{fabi01}.  
However, these scenarios, as clever as they are, may require fine-tuning 
and are otherwise difficult to prove observationally.
In any case, they leave the question
of the repository for most of the cooling gas unanswered.

\begin{deluxetable*}{clccccc}
\tabletypesize{\scriptsize}
\tablecaption{Radio properties. \label{tbl-cluster-radio}}
\tablewidth{400pt}
\tablehead{
	\colhead{} & \colhead{} & \colhead{$\sigma$ (ref)\tablenotemark{a}} & \colhead{$S_{1400}$}   &
	\colhead{} & \colhead{$P_{1400}$} & \colhead{$L_{\rm{radio}}$} \\
	\colhead{System\phn} & \colhead{z} & \colhead{(km s$^{-1}$)} & \colhead{(Jy)} &
	\colhead{$\alpha$ (ref)\tablenotemark{b}} &	\colhead{($10^{24}$ W Hz$^{-1}$)} & \colhead{($10^{42}$ ergs s$^{-1}$)}
}
\startdata
    Cygnus A    & 0.056  & \nodata         & 1598$\pm$41\phn\phn       & 0.7\phm{2} (4)\phm{22222,,,}         & 11800$\pm$300\phn\phn     & 700$\pm$20\phn       \\
    Hydra A     & 0.052  & \phn \phn \phn322$\pm$20 (8)\phn  & 40.8$\pm$1.3\phn      & 0.92 (4,13)\phm{222,,}     & 261$\pm$8\phn\phn         & 19.9$\pm$0.6\phn     \\
	A2597       & 0.085  & \phn \phn \phn224$\pm$19 (16) & 1.875$\pm$0.056   & 1.35 (11)\phm{2222,,,}       & 35$\pm$1\phn          & 6.8$\pm$0.2         \\
	MKW 3S      & 0.045  & \nodata         & 115.0$\pm$3.9\phn\phn     & 2.3\phm{2} (4,13,17)\phm{2,}   & 0.58$\pm$0.02     & 3.6$\pm$0.1        \\
	A2052       & 0.035  & \phn \phn \phn253$\pm$12 (19) & 5.50$\pm$0.21     & 1.2\phn (1,4,13)\phm{22,}    & 15.7$\pm$0.6\phn      & 2.08$\pm$0.08       \\
	A133        & 0.060  & \nodata         & 0.167$\pm$0.006   & 1.9\phn (14,15)\phm{22,,}     & 1.54$\pm$0.06     & 1.94$\pm$0.07       \\
	A4059       & 0.048  & \phn \phn \phn296$\pm$49 (6)\phn  & 1.284$\pm$0.043   & 1.43 (13,18,21)\phm{,} & 6.96$\pm$0.2\phn      & 1.73$\pm$0.06       \\
	A2199       & 0.030  & \phn \phn \phn295$\pm$6\phm{9} (7)\phn   & 3.58$\pm$0.12     & 1.37 (1,4,17)\phm{22,}   & 7.43$\pm$0.2\phn      & 1.56$\pm$0.05       \\
	Perseus     & 0.018  & \phn \phn \phn246$\pm$10 (8)\phn  & 22.83$\pm$0.68\phn    & 1.0\phn (10)\phm{2222,,,}        & 16.5$\pm$0.5\phn      & 1.42$\pm$0.04      \\ 
	RBS 797     & 0.350  & \nodata         & 0.0217$\pm$0.0008 & \nodata         & 9.0$\pm$0.3       & 0.78$\pm$0.03     \\  
	A1795       & 0.063  & \phn \phn \phn297$\pm$12 (3)\phn  & 0.925$\pm$0.028   & 0.98 (1,4,13)\phm{22,}   & 9.0$\pm$0.3       & 0.75$\pm$0.02      \\ 
	M87         & 0.0042 & \phn \phn \phn330$\pm$5\phm{2} (2)\phn   & 138.5$\pm$4.9\phn\phn     & 0.81 (4,9,13)\phm{22,}   & 5.48$\pm$0.2\phn      & 0.36$\pm$0.01       \\
	Centaurus   & 0.011  & \phn \phn \phn256$\pm$11 (5)\phn  & 3.8            & 0.7\phn (4,9,14,20) & 1.1            & 0.060            \\	
	A478        & 0.081  & \nodata         & 0.0369$\pm$0.0015 & \nodata         & 0.60$\pm$0.02     & 0.052$\pm$0.002    \\ 
	M84         & 0.0035 & \phn \phn \phn278$\pm$4\phm{2} (8)\phn  & 6.00$\pm$0.15     & 0.63  (4,9,17)\phm{22,}  & 0.162$\pm$0.004   & 0.0089$\pm$0.0002   \\
    2A 0335+096 & 0.035  & \nodata         & 0.0367$\pm$0.0018 & 0.9\phn (12)\phm{2222,,,}        & 0.104$\pm$0.005   & 0.0077$\pm$0.0004 \\
	A262        & 0.016  & \nodata         & 0.0657$\pm$0.0023 & 0.6\phn (4)\phm{22222,,,}         & 0.039$\pm$0.001   & 0.00210$\pm$0.00007 \\
	HCG 62      & 0.014  & \nodata         & 0.0049$\pm$0.0005 & \nodata         & 0.0021$\pm$0.0002 & 0.00018$\pm$0.00002 \\
\enddata

\tablenotetext{a}{When no velocity dispersion was available, $\left< \sigma \right> = 280$ km s$^{-1}$ was adopted. References are in parentheses.}
\tablenotetext{b}{The spectral index is defined so that $S \sim \nu^{-\alpha}$. When no spectral index was available, $\alpha = 1$ was adopted. References are in parentheses.}
\tablerefs{
	(1) Becker, White, \& Edwards 1991; (2) Bender, Saglia, \& Gerhard 1994; (3) Blakeslee \& Tonry 1992; (4) Burbidge \& Crowne 1979; (5) Carollo, Danzinger, \& Buson 1993; (6) Carter et al. 1985; (7) Fisher, Illingworth, \& Franx 1995; (8) Heckman et al. 1985; (9) Kuehr et al. 1981; (10) Pedlar et al. 1990; (11) Sarazin et al. 1995; (12) Sarazin, Baum, \& O'Dea 1995; (13) Slee 1995; (14) Slee \& Siegman 1988; (15) Slee et al. 2001; (16) Smith, Heckman, \& Illingworth 1990; (17) Spindrad et al. 1985; (18) Taylor, Barton, \& Ge 1994; (19) Tonry 1985; (20) Wright et al. 1994; (21) Wright et al. 1996}
\end{deluxetable*}

The more appealing interpretation, which we address in this paper, 
posits that radiation losses are being balanced, or nearly so,
by heating, implying that the large cooling rates of the last
decade were indeed overestimated.   This suggestion has its own difficulties.
Maintaining gas with a cooling time approaching 
100 Myr at keV temperatures almost certainly requires
the existence of one or more heating mechanisms operating
in a self-regulating feedback loop. 
One such mechanism, thermal conduction from 
the hot outer layers of clusters, may be energetically
feasible, in some instances \citep[e.g.\
][]{tuck83,bert86,zaka03,voig02,voig04}.
However, it generally
requires fine tuning and can be unstable \citep{bregdav,soke03}.
Moreover, conduction operating alone at even the Spitzer rate
cannot offset radiation losses \citep{voig02,wise03} in all clusters,
and is therefore unlikely to provide a general solution 
to the heating problem 
(recent simulations by Dolag et al.\ 2004 support this conclusion). Additional
heat sources, such as cosmic rays \citep{boeh88,loew91}, and supernova
explosions \citep{mcna04} may contribute to
heating the gas.  Nevertheless, these mechanisms are generally 
incapable of balancing radiation losses.  

In this paper,  we evaluate whether 
the mechanical energy generated by active 
galactic nuclei (AGN) can balance radiation losses
in cluster cores.  This possibility, which has a substantial legacy 
in the literature \citep[e.g.\ ][]{tabo93,binn95,tuck97,ciot01,sok01}, 
has been rejuvenated by the crisp, new {\it Chandra} images of clusters  
showing the keV gas being displaced by radio sources harbored by central
cluster galaxies.  The now ubiquitous signature of
these interactions are X-ray surface brightness depressions 
projected on the radio lobe emission at 1.4 GHz, as is seen in Perseus
\citep{boeh93,fabi00,schm02,fabi02b,fabi03a,fabi03b},
Cygnus A \citep{carilli}, and Hydra~A 
\citep{mcna00,davi01,nuls02}. The displacement of the gas
creates a low-density, rising bubble in 
pressure balance with the surrounding medium. 
X-ray surface brightness
depressions that have no obvious association with the bright radio emission
at 1.4 GHz, the so-called ghost cavities,
have also been found, such as those 
in Abell~2597 \citep{mcna01} and the outer depressions 
in Perseus \citep{fabi00}.  These depressions 
are thought to have been created by interactions that occurred
in the more distant past, but whose radio emission has faded over time.

This general scenario has been modeled theoretically
using a variety of hydrodynamical, magnetohydrodynamical,
and analytical techniques, which have successfully reproduced 
the gross characteristics of the cavities 
\citep[e.g.\ ][]{gull73,chur01,brug01,quil01,brig02,brug02b,reyn02,brug03,ddy03,bass03,binn03,kais03,math03,omma04,robi04}. 
The models do not, however, predict with any certainty
the amount of mechanical energy provided by radio sources of a given
luminosity, nor their frequency of recurrence.    
Whereas the cavities in several individual 
objects, such as Perseus \citep{fabi03a} and Hydra A \citep{mcna00},
contain enough enthalpy to balance cooling, at least for a short
period of time, a
systematic survey of cavities in systems with a broad range of properties
is required to  determine whether this is generally true.
In this paper we address 
this question by setting observational limits on the energetics
and ages of cavities in 18 systems taken 
from the \textit{Chandra} Data Archive.  We adopt 
$H_{0} = 70~\mbox{km s$^{-1}$ Mpc$^{-1}$}$, $\Omega_{M} = 0.3$, 
and $\Omega_{\Lambda} = 0.7$ in all calculations throughout this paper.

\section{The Sample}\label{sec-sample}

Approximately 80 systems from the \textit{Chandra} Data Archive were visually inspected for 
surface brightness depressions.  Of these, we selected the 18 systems
having well defined surface brightness depressions
associated with their radio sources (see Table~\ref{tbl-cluster-radio}).  Of the 18 systems, 16 were imaged with the ACIS-S3 detector and two
with the ACIS-I3 detector (RBS~797 and MKW~3S), with exposure 
times ranging from 12~ksec (RBS~797) to 50~ksec (HCG~62).  The sample consists of 16 
galaxy clusters, one galaxy group (HCG~62), and one giant elliptical galaxy (M84), ranging in redshift
from $z=0.0035$ (M84) to $z=0.35$ (RBS~797), and in X-ray luminosity from 
$\sim 10^{41}~\mbox{ergs s$^{-1}$}$ 
(M84) to $4\times 10^{45}~\mbox{ergs s$^{-1}$}$ (RBS~797).  
We avoided depressions with questionable association with a radio 
source, and we excluded clusters in which 
there is clear evidence for merging, since merging clusters often 
show complex structure that can be mistaken for a radio-induced 
cavity.  All clusters in our sample were previously reported in the 
literature as containing cavities likely to be associated with radio bubbles.

\section{Data Reduction and Analysis}

\subsection{Radio Analysis}

The systems in our sample have a wide range of radio properties, from
powerful double-lobed FR~II radio sources with luminosities of
$\sim 7\times 10^{44}~\mbox{ergs sec$^{-1}$}$ (Cygnus~A) to weak sources
with luminosities of $\sim 2\times 10^{38}~\mbox{ergs sec$^{-1}$}$ (HCG~62).  
Table~\ref{tbl-cluster-radio} gives the radio properties of our sample.  
The radio power at $\nu = 1400~\mbox{MHz}$ was calculated as 
$P_{\nu} = 4\pi D_{\nu}^{2} S_{\nu},$ where 
$D_{\nu} = D_{L} (1+z)^{-(1+\alpha)/2}$ \citep{vonh74}.
The total radio luminosity was calculated by integrating the flux between
$\nu_{1} = 10~\mbox{MHz}$ and $\nu_{2} = 5000~\mbox{MHz}$ as
\begin{equation}
L_{\rm{rad}} = 4\pi D_{L}^{2} S_{\nu_{0}} \int_{\nu_{1}}^{\nu_{2}} 
\left( \nu / \nu_{0} \right)^{-\alpha} \, d\nu ,
\end{equation}
where we have assumed a power law spectrum ($S_{\nu} \sim \nu^{-\alpha} ,$ where $\alpha$ is the
spectral index).  We used for $S_{\nu_{0}}$ the 1400~MHz flux from the NRAO
VLA\footnote{The VLA (Very Large Arry) is a facility of the National Radio Astronomy Observatory (NRAO).  
The NRAO is a facility of the National Science Foundation operated under cooperative 
agreement by Associated Universities, Inc.} Sky Survey (NVSS) catalog \citep{cond98}, except in the case of the 
Centaurus cluster where no NVSS data were available.  In this case, we used the
1410~MHz flux from the Parkes Radio Sources Catalogue \citep{wrig90}.  Spectral 
indices were taken from the catalogs referenced in Table~\ref{tbl-cluster-radio}.  
As the derived spectral index can vary depending on the frequencies used, 
we have adopted a weighted average of the 
available spectral indices.  In cases in which the spectral index was not 
available, a value of $\alpha = 1$ was adopted.

\phn

\subsection{X-ray Analysis}

The X-ray data were obtained through the \textit{Chandra} Data Archive 
and were reprocessed with CIAO, version 2.3, using CALDB, version 2.21.  The charge transfer 
inefficiency (CTI) correction was applied during reprocessing of the 
level 1 event file.  Blank-sky background files were used for background 
subtraction for all clusters.\footnote{See http://asc.harvard.edu/contrib/maxim/acisbg/}  
The background files were normalized to the count rate of the 
source image in the 10-12~keV band, after the removal of all bright emission.  The required adjustment
was less than 12\% for all clusters except Centaurus and Perseus,
which both required background adjustments of $\sim 30\%$.  Spectra with at
least 2000 counts were extracted in circular annuli centered on the
X-ray centroid of the cluster.  Response files
were made using the CIAO tools \emph{mkrmf} and \emph{mkwarf}.  
We attempted to correct the resulting ARFs for the quantum 
efficiency degradation problem using the \emph{corrarf} 
tool.\footnote{See http://asc.harvard.edu/cal/Acis/Cal\_prods/qeDeg/}  
However, upon spectral fitting, we found that $\sim 75\%$ 
of our sample was overcorrected by \emph{corrarf}.  Therefore,  we present our 
results without the correction applied \citep[for a discussion of this problem, see][]{voig04}.  
In general, the largest effect of using \emph{corrarf} was on the cooling rates, 
which increased on average by a factor of 2 after the correction was applied.

To find radial temperatures and densities, we deprojected the 
spectra extracted above.  The deprojection was 
performed assuming spherical symmetry and
using the PROJCT model in XSPEC 11.2.0 with a single-temperature plasma model
(MEKAL) and foreground absorption (WABS), fitted between energies of 0.5 
and 7.0~keV.  The foreground column density, $N_{\rm{H}},$ 
was tied between annuli and allowed to vary. The MEKAL abundance was also free to vary.  
The redshift was fixed to the value given in Table~\ref{tbl-cluster-radio}.  
The density was then calculated from the normalization of the MEKAL component, 
assuming $n_{\rm{e}}=1.2n_{\rm{H}}$ (for a fully ionized gas with hydrogen and helium mass fractions of $X=0.7$ 
and $Y=0.28$).  The pressure in each annulus was calculated as $P=nkT,$ 
where we have assumed an ideal gas and $n \approx 2n_{\rm{e}}$.

The luminosity of the cooling gas inside the cooling radius is needed
to investigate whether or not AGN heating can balance cooling. 
Within the cooling radius, radiative energy losses must
be replaced to prevent the deposition of large quantities of cool gas. 
We define the cooling radius as the radius 
within which the gas has a cooling time less than $7.7\times 10^{9}~\mbox{yr}$, 
the look-back time to $z=1$ for our adopted cosmology.  This redshift
is roughly the distance to which clusters have been found with properties
similar to present day clusters.  The corresponding lookback time
should then approximate the time a cooling flow has had to establish itself.
The cooling time was calculated using the cooling curves of \citet{boeh89}.  
Table~\ref{tbl-cluster-xray} gives the values of $r_{\rm{cool}}$ for each cluster.   
Within this radius, we performed deprojections by fitting both a 
cooling model and a single temperature model to the spectra. 

In order to obtain a spectroscopic estimate of the cooling luminosity, 
we performed the deprojection using a cooling flow model 
(PROJCT*WABS*[MEKAL+MKCFLOW]), fit between 0.5 and 7.0~keV.
To force all cooling to be within the cooling radius, the MKCFLOW model was 
used only inside the cooling radius and was set to zero outside.
The MKCFLOW low temperature was fixed to 0.1~keV, resulting in
an estimate of the luminosity of gas cooling to low temperatures.  
Within each annulus, the MEKAL and MKCFLOW abundances were tied together, 
and the MKCFLOW high temperature was tied to the temperature of the MEKAL 
component.  Lastly, the column density ($N_{\rm{H}}$) was tied between annuli and allowed to vary.  
The spectroscopic estimate of the bolometric cooling luminosity
inside the cooling radius, $L_{\rm{spec}},$ was then calculated
from the unabsorbed fluxes obtained from the MKCFLOW model, integrated
between energies of $\sim 0.1-100~\mbox{kev}$.

To find the total luminosity inside the cooling radius, we  
performed the deprojection using a single-temperature 
model (PROJCT*WABS*MEKAL) fit between 0.5 and 7.0~keV to the 
same spectra used with the cooling flow model, again 
with $N_{\rm{H}}$ tied between regions and allowed to vary.  
The unabsorbed fluxes from the MEKAL components for the 
annuli within the cooling radius, extrapolated between 
$\sim 0.1-100~\mbox{kev},$ were used to find the
bolometric luminosity of the X-ray emitting gas, $L_{\rm{X}}$.  
Table~\ref{tbl-cluster-xray} gives $L_{\rm{X}}$ 
and $L_{\rm{spec}}$ (with $1 \sigma$ errors estimated by XSPEC) 
for each object in our sample.  
Typically, $L_{\rm{spec}}$ is approximately 10\% of $L_{\rm{X}}$ for our sample.
Our values for $L_{\rm{spec}}$ and $L_{\rm{X}}$ are in 
reasonable agreement with published values for most of our sample.

\begin{deluxetable*}{ccccccc}
\tabletypesize{\scriptsize}
\tablecaption{Results of X-ray spectral modeling. \label{tbl-cluster-xray}}
\tablewidth{350pt}
\tablehead{
	\colhead{} & \colhead{} & \multicolumn{2}{c}{PROJCT*WABS*MEKAL} & \multicolumn{3}{c}{PROJCT*WABS*(MEKAL+MKCFLOW)} \\
 	\colhead{} & \colhead{$r_{\rm{cool}}$} & \colhead{} & \colhead{$L_{\rm{X}}(<r_{\rm{cool}})$} & \colhead{} & \colhead{$\dot{M}$} & \colhead{$L_{\rm{spec}}(<r_{\rm{cool}})$} \\
	\colhead{System\phn} & \colhead{(kpc)} & \colhead{$\chi^{2}$/dof} & \colhead{($10^{42}$ ergs s$^{-1}$)} & \colhead{$\chi^{2}$/dof} & \colhead{$\rm{M_{\odot}}$ yr$^{-1}$} & \colhead{($10^{42}$ ergs s$^{-1}$)}
}
\startdata
	RBS 797     & 191    & 154/183    & $4500_{-700}^{+800}$ & 151/180    & $880_{-670}^{+1800}$ & $1200_{-1100}^{+4600}$  \\
	A478        & 150    & 3704/2796  & 1220$\pm$60          & 3671/2792  & $150_{-68}^{+60}$    & $180_{-95}^{+60}$   \\
	Perseus     & 102    & 9348/5717  & $670_{-30}^{+40}$    & 9165/5703  & $54_{-18}^{+48}$     & $59_{-17}^{+31}$  \\
	A1795       & 137    & 2476/1721  & 490$\pm$30           & 2487/1718  & $18_{-10}^{+12}$     & $11_{-5}^{+9}$ \\
	A2597       & 129    & 1341/1149  & $430_{-30}^{+40}$    & 1329/1145  & $59 \pm 40$          & 28$\pm$19    \\
    Cygnus A    & \phn78     & 3062/2234  & 410$\pm$30           & 3062/2230  & $<8$                 & $<6$    \\
    2A 0335+096 & 122    & 2439/1903  & 290$\pm$20           & 2357/1897  & $120 \pm 30$         & $44_{-11}^{+12}$ \\
    Hydra A     & 100    & 1658/1486  & 250$\pm$15           & 1652/1482  & $14_{-7}^{+9}$       & $8.1_{-3.7}^{+4.8}$ \\
	A2199       & 113    & 2422/1971  & 150$\pm$10           & 2421/1966  & $2.0_{-1.9}^{+7.0}$  & $1.2_{-1.1}^{+4.5}$ \\
	A133        & \phn92     & 1624/1160  & 95$\pm$5             & 1598/1158  & $25 \pm 6$           & 11$\pm$3 \\
	MKW 3S      & \phn88     & 2242/2037  & 92$\pm$7             & 2242/2033  & $<2$                 & $<1$           \\
	A2052       & 101    & 2934/2035  & $84_{-5}^{+6}$       & 2867/2030  & $12_{-3}^{+4}$     & $4.4_{-1.2}^{+1.7}$ \\
	A4059       & \phn99     & 1396/1109  & $71_{-7}^{+8}$       & 1296/988\phn   & $6.7_{-4.1}^{+8.5}$  & $4.7_{-3.3}^{+3.9}$ \\
	Centaurus   & \phn62     & 5100/2177  & 30$\pm$2             & 4932/2171  & $10.2_{-1.2}^{+1.5}$ & $2.8_{-0.4}^{+0.6}$ \\
	A262        & \phn66     & 1978/1169  & 13.8$\pm$1.0         & 1923/1165  & $1.5_{-0.4}^{+0.7}$  & $1.5_{-0.4}^{+0.7}$  \\
	M87         & \phn35     & 15540/5714\phn & $9.8_{-0.7}^{+0.8}$  & 15622/5688\phn & $1.8_{-0.6}^{+1.2}$  & $0.62_{-0.18}^{+0.34}$ \\
	HCG 62      & \phn33     & 1008/626\phn   & 2.0$\pm$0.2          & 961/624    & $1.1_{-0.2}^{+0.3}$  & $0.20_{-0.03}^{+0.04}$ \\
	M84         & \phn14     & 682/450    & 0.09$\pm$0.01        & 701/448    & $0.06 \pm 0.05$      & $0.017_{-0.004}^{+0.016}$ \\
\enddata
\end{deluxetable*}

\phn

\phn

\phn

\section{X-ray Surface Brightness Depressions}

In total, 36 surface brightness depressions or cavities were identified in the 18 systems.
Table~\ref{tbl-bubbles} lists the cavity properties.  
The cavities are classified as either radio-filled or radio-faint
ghosts, depending on the presence of 1400~MHz or higher 
frequency radio emission inside the cavity.  The cavities 
are classified as 
radio-filled if there is a 
direct anticorrelation between the radio emission at 
1400~MHz or higher and the X-ray emission, 
such that the radio emission fills preferentially the X-ray surface brightness depressions.  
The cavities classified as ghost cavities, while possibly possessing significant 
radio emission at frequencies at or below 1400~MHz, do not show the
anti-correlation between the 
high-frequency radio emission and the X-ray emission.  Our 
classification scheme relies heavily on the availability of
high-resolution radio images at several frequencies.  
However, the radio data available are 
inhomogeneous, and classifying the objects lacking high -resolution radio images was challenging.  The 
poor radio images available for A262, RBS~797, and 
HCG~62, in particular, led us to classify 
them as ghosts (see E.\ L.\ Blanton et al.\ 2004, in preperation). 

For each cavity, a size and position were measured, assuming the cavity 
extends to the inner edge of any bright surrounding emission.  
The projected shapes of the cavities were measured by eye as circles 
or ellipses from the exposure-corrected, unsmoothed images.  This is 
a qualitative measurement, the accuracy of which depends on the 
signal-to-noise ratio of the image and on the contrast of the cavity
with its surroundings.  To distinguish between the poorly defined 
and well-defined cavities, we have assigned a figure of merit (FOM) to 
each cavity ranging from 1 for the best-defined 
cavities -- those with surrounding bright rims -- to 3 for the worst-defined ones without bright rims.  

In the analysis that follows, we assumed that the cavities are bubbles devoid
of gas at the local ambient temperature \citep{mcna00, blan03}.
Their volumes were calculated assuming spherical or
prolate ellipsoidal shapes, 
with semimajor axis $a$ and semiminor axis $b$.  The errors 
in the volumes due to projection were estimated by 
allowing each bubble to have an intrinsic $a/b$ as large as that 
of the most eccentric cavity observed in the sample, 
$\left( a/b \right)_{\rm{max}}$.  The upper and lower 
limits are calculated assuming either oblate or prolate symmetry.  
In this sense, spherical bubbles have the greatest range of possible 
volumes, while projected ellipses with an 
$a/b =\left( a/b \right)_{\rm{max}}^{2/3}$ have the smallest range.  
The pressure and temperature of the gas surrounding the cavity 
were taken to be the azimuthally averaged values at the projected radius
of its center.  The work done on the surrounding 
medium by the cavity is then simply $W_{\rm{bub}} = pV$, if it
expands slowly compared to the sound speed.

\phn

\phn

\subsection{Cavity Ages}\label{sec-cavage}

\begin{deluxetable*}{cccccccccc}
\tabletypesize{\scriptsize}
\tablecaption{Cavity properties. \label{tbl-bubbles}}
\tablewidth{400pt}
\tablehead{
	\colhead{} & \colhead{Cavity} & \colhead{$a$\tablenotemark{b}}   &
	\colhead{$b$\tablenotemark{c}} & \colhead{$R$\tablenotemark{d}}  & \colhead{$p V$\tablenotemark{e}} & 
	\colhead{$t_{\rm{c_{s}}}$} & \colhead{$t_{\rm{buoy}}$} & \colhead{$t_{\rm{r}}$} & \colhead{} \\
	\colhead{System\phn} & \colhead{Type (FOM)\tablenotemark{a}} & \colhead{(kpc)} &
	\colhead{(kpc)} & \colhead{(kpc)} & \colhead{($10^{57}$ergs)} & 
	\colhead{($10^{7}$ yr)} & \colhead{($10^{7}$ yr)} & \colhead{($10^{7}$ yr)} & \colhead{ref}
}
\startdata
	RBS 797     & G (2) & \phn9.7  & \phn9.7  & 19.5 & $190_{-120}^{+340}$          & 1.4  & \phn3.6  & \phn6.8 & 18      \\
	            & G (2) & 13.4 & 8.5  & 23.8 & $190_{-40}^{+150}$           & 1.7  & \phn5.2  & \phn7.8 &         \\
	A478        & F (2) & \phn5.5  & \phn3.4  & \phn9.0  & $5.8_{-1.1}^{+4.4}$          & 1.1  & \phn1.9  & \phn3.1 & 20      \\
	            & F (2) & \phn5.6  & \phn3.4  & \phn9.0  & $6.0_{-1.1}^{+4.3}$          & 1.1  & \phn1.9  & \phn3.1 &         \\
	A1795       & G (3) & 18.5 & \phn7.2  & 18.5 & $39_{-4}^{+53}$              & 1.8  & \phn3.7  & \phn6.8 & 5       \\
	Perseus     & F (1) & \phn4.7  & \phn4.7  & \phn7.0  & $12_{-8}^{+21}$              & 0.6  & \phn1.3  & \phn3.2 & 6,7     \\
	            & F (1) & \phn7.3  & \phn6.3  & 11.2 & $24_{-13}^{+34}$             & 1.0  & \phn2.1  & \phn4.8 &         \\
	            & G (1) & 12.9 & \phn6.3  & 29.7 & $24_{-1}^{+12}$              & 3.0  & \phn9.4  & \phn9.1 &         \\
	            & G (2) & 13.6 & \phn4.8  & 36.7 & $14_{-1}^{+25}$              & 3.6  & 15.1 & \phn9.7 &         \\
	Cygnus A    & F (1) & 29.0 & 17.2 & 43.0 & $470_{-30}^{+300}$           & 3.2  & \phn8.9  & 15.3& 19      \\
	            & F (1) & 33.8 & 23.1 & 44.7 & $930_{-230}^{+860}$          & 3.3  & \phn8.1  & 17.4&         \\
	A2597       & G (2) & 10.2 & \phn7.1  & 22.6 & $21_{-5}^{+20}$              & 2.6  & \phn6.6  & \phn8.6 & 15      \\
	            & G (2) & \phn7.1  & \phn7.1  & 23.1 & $14_{-9}^{+26}$              & 2.6  & \phn6.8  & \phn7.9 &         \\
	2A 0335+096 & G (2) & \phn9.3  & \phn6.5  & 23.1 & $9.7_{-2.8}^{+9.6}$          & 3.2  & \phn5.7  & \phn6.6 & 13      \\
	            & G (3) & \phn4.8  & \phn2.6  & 27.5 & $0.79_{-0.02}^{+0.40}$       & 3.9  & 11.7 & \phn4.8 &         \\
	Hydra A     & F (2) & 17.7 & 11.8 & 28.5 & $83_{-17}^{+73}$             & 3.0  & \phn5.1  & \phn8.7 & 4,14,16 \\
	            & F (2) & 19.9 & 11.7 & 33.8 & $87_{-7}^{+57}$              & 3.2  & \phn5.6  & \phn9.3 &         \\
	A2199       & F (2) & \phn6.5  & \phn6.5  & 18.9 & $6.6_{-4.3}^{+11.9}$         & 2.1  & \phn4.0  & \phn5.2 & 11      \\
	            & F (2) & \phn6.2  & \phn3.5  & 21.2 & $1.8_{-0.1}^{+1.1}$          & 2.2  & \phn6.5  & \phn4.7 &         \\
	MKW 3S      & G (3) & 53.9 & 22.6 & 58.7 & $300_{-14}^{+310}$           & 5.8  & 12.4 & 22.3& 12      \\
	A2052       & F (1) & \phn6.5  & \phn6.0  & \phn6.7  & $3.2_{-1.8}^{+5.0}$          & 1.1  & \phn1.0  & \phn3.5 & 1,2     \\
	            & F (1) & 10.7 & \phn7.8  & 11.2 & $8.4_{-2.8}^{+8.8}$          & 1.8  & \phn2.0  & \phn5.5 &         \\
	A4059       & G (2) & \phn9.2  & \phn9.2  & 19.3 & $8.9_{-6.0}^{+16.3}$         & 2.8  & \phn3.5  & \phn6.2 & 10      \\
	            & G (2) & 20.4 & 10.1 & 22.7 & $24_{-3}^{+11}$              & 2.6  & \phn4.2  & \phn8.4 &         \\
	A133        & F (2) & \phn9.8  & \phn5.2  & 28.1 & $3.4_{-0.1}^{+1.7}$          & 3.5  & \phn8.6  & \phn7.0 & 9       \\
	            & F (2) & \phn9.8  & \phn5.7  & 32.7 & $4.0_{-0.1}^{+2.7}$          & 3.9  & 10.2 & \phn7.7 &         \\
	Centaurus   & F (3) & \phn3.3  & \phn1.6  & \phn3.5  & $0.196_{-0.003}^{+0.10}$     & 0.7  & \phn0.7  & \phn1.5 & 17      \\
	            & F (2) & \phn3.3  & \phn2.4  & \phn6.0  & $0.40_{-0.13}^{+0.42}$       & 1.0  & \phn1.3  & \phn2.2 &         \\
	A262        & G (2) & \phn2.6  & \phn2.6  & \phn6.2  & $0.17_{-0.11}^{+0.32}$       & 1.1  & \phn1.3  & \phn2.0 & 3       \\
	            & G (3) & \phn3.3  & \phn2.6  & \phn6.7  & $0.21_{-0.10}^{+0.27}$       & 1.2  & \phn1.4  & \phn2.2 &         \\
	M87         & F (2) & \phn1.6  & \phn0.8  & \phn2.2  & $0.078_{-0.003}^{+0.043}$    & 0.4  & \phn0.4  & \phn0.6 & 22      \\
	            & F (2) & \phn2.3  & \phn1.4  & \phn2.8  & $0.25_{-0.02}^{+0.18}$       & 0.4  & \phn0.4  & \phn0.9 &         \\
	HCG 62      & G (2) & \phn5.0  & \phn4.3  & \phn8.4  & $0.29_{-0.15}^{+0.41}$       & 1.8  & \phn1.5  & \phn3.1 & 21      \\
	            & G (2) & \phn4.0  & \phn4.0  & \phn8.6  & $0.21_{-0.13}^{+0.37}$       & 1.9  & \phn1.6  & \phn2.9 &         \\
	M84         & F (2) & \phn1.6  & \phn1.6  & \phn2.3  & $0.019_{-0.013}^{+0.035}$    & 0.5  & \phn0.4  & \phn1.0 & 8       \\
	            & F (2) & \phn2.1  & \phn1.2  & \phn2.5  & $0.013_{-0.001}^{+0.007}$    & 0.6  & \phn0.5  & \phn1.0 &         \\
\enddata
\tablenotetext{a}{Radio-filled cavities are denoted by ``F,'' radio-faint ghosts are denoted by ``G.'' The FOM gives a relative measure of the cavity's contrast to its surroundings: (1) high contrast: bright rim surrounds cavity; (2) medium contrast: bright rim partially surrounds cavity; and (3) low contrast: no rim or faint rim surrounds cavity.}
\tablenotetext{b}{Projected semimajor axis of the cavity.}
\tablenotetext{c}{Projected semiminor axis of the cavity.}
\tablenotetext{d}{Projected distance from the cavity center to the radio core.}
\tablenotetext{e}{The errors in $pV$ include an estimate of the projection effects; see the text for details.}
\tablerefs{
	(1) Blanton et al. 2001; (2) Blanton, Sarazin, \& McNamara 2003; (3) Blanton et al. 2004; (4) David et al. 2001; (5) Ettori et al. 2002; (6) Fabian et al. 2000a; (7) Fabian et al. 2002; (8) Finoguenov \& Jones 2001 (9) Fujita et al. 2002; (10) Heinz et al. 2002; (11) Johnstone et al. 2002; (12) Mazzotta et al. 2002; (13) Mazzotta, Edge, \& Markevitch 2003; (14) McNamara et al. 2000; (15) McNamara et al. 2001; (16) Nulsen et al. 2002; (17) Sanders \& Fabian 2002; (18) Schindler et al. 2001; (19) Smith et al. 2002; (20) Sun et al. 2003; (21) Vrtilek et al. 2002; (22) Young, Wilson, \& Mundell 2002}
\end{deluxetable*}

The age of each cavity was calculated in three ways. 
First, as the time required for the cavity 
to rise the projected distance from the radio 
core to its present location at the speed of sound, 
$v_{\rm{c_{s}}} = (\gamma kT/\mu m_{\rm{H}})^{1/2},$
where we have taken $\gamma = 5/3$ and $\mu = 0.62$.  The cavity age is then
\begin{equation}
t_{\rm{c_{s}}} = R/v_{\rm{c_{s}}}=R\sqrt{\mu m_{\rm{H}}/\gamma kT},
\end{equation}
where $R$ is the projected distance from the center of the bubble to the radio core.  This scenario is 
favored in the computational modeling of \citet{omma04}, in which the 
bubble is produced by a high-momentum jet from the AGN instead of
rising buoyantly.   
Second, the age was calculated as the time required for the cavity to 
rise buoyantly (bubble-like) at its terminal velocity  
$v_{\rm{t}} \sim (2 g V / S C)^{1/2},$
where $V$ is the volume of the bubble, $S$ is the cross section of the bubble, 
and $C = 0.75$ is the drag coefficient \citep{chur01}.  
The gravitational acceleration was calculated using the 
stellar velocity dispersion of the central galaxy, under the approximation that the galaxy is an isothermal sphere,
as $g \approx 2 \sigma^{2} / R$ \citep{binn87}. Published values of the velocity dispersion were 
used when available (see Table~\ref{tbl-cluster-radio}); otherwise, the average 
value ($\left< \sigma \right> = 280~\mbox{km s$^{-1}$}$) was adopted.  
The cavity age is then given by 
\begin{equation}
t_{\rm{buoy}} = R/v_{\rm{t}} \sim R\sqrt{SC/2gV}.
\end{equation}
Finally, the age was calculated as the time required to refill the displaced
volume as the bubble rises upward \citep{mcna00,nuls02}:
\begin{equation}
t_{\rm{r}} \sim 2 R \sqrt{r / G M(R)} = 2\sqrt{r/g},
\end{equation}
where $r$ is the radius of the cavity 
[for ellipsoidal cavities, $r = (a b)^{1/2}$].

In general, the ages calculated using the speed of sound are the shortest, 
those based on the refilling time scale are the greatest, and those 
calculated using the terminal velocity lie in between.
The instantaneous  mechanical luminosity per cavity or cavity pair
is then  $L_{\rm{mech}} = W_{\rm{bub}} / t,$ where $t$ is 
the age of the bubble.  

\begin{figure*}
\plottwo{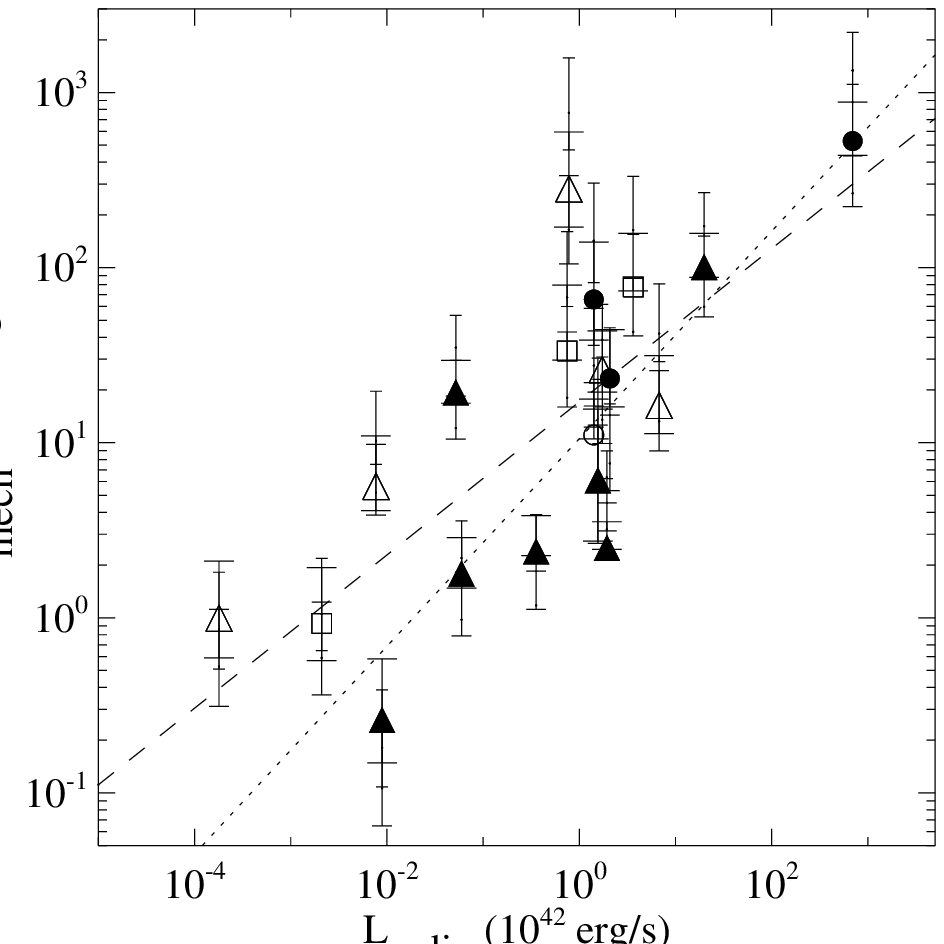}{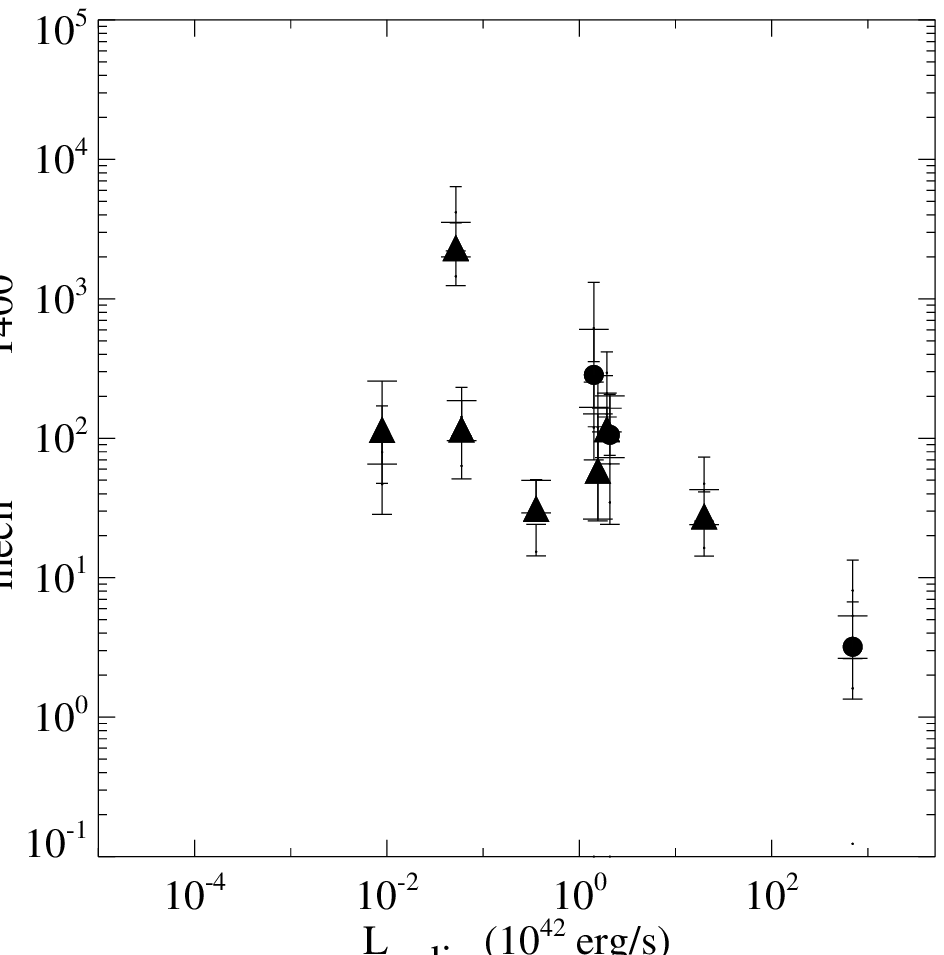}
\caption{\emph{Left:} Mechanical luminosity vs.\ radio luminosity.
The symbols and wide error bars denote the values of the 
mechanical luminosity calculated using the buoyancy timescale.
The short and medium-width error bars denote upper and lower limits of the 
mechanical luminosity calculated using the sound 
speed and refill timescales, respectively.
The different symbols indicate FOMs of (\emph{circles}), 2 (\emph{triangles}), 
and 3 (\emph{squares}).  Filled symbols denote radio-filled 
cavities, and empty symbols denote ghost cavities.  
Each point represents the sum of mechanical luminosities 
of each bubble type.  The best-fit lines are shown for the entire 
sample (\emph{dashed line}) and for the radio-filled cavities only 
(\emph{dotted line}).  \emph{Right:} The ratio of mechanical 
luminosity to $\nu P_{\nu}$ at 1400~MHz  vs. radio luminosity 
for the radio-filled cavities only. \label{fig-Lr_Lm}}
\end{figure*}

This approximation to the mechanical
luminosity of the rising bubbles is highly uncertain.
In calculating $L_{\rm{mech}}$, a measurement of the average timescale for
the radio source to replenish itself is required. This timescale
is unknown for each source, but perhaps this 
can best be measured in objects with clearly defined
ghost cavities and a detached central source,  where the duty cycle is clearly
evident.  This situation has been noted in two objects, 
Perseus \citep{fabi00} and A2597 \citep{mcna01} whose ghost cavities range in age
between $\sim 5-8\times 10^7~{\rm yr}$, with a likely age of
$\sim 10^8~{\rm yr}$ when projection is taken into account.  Furthermore,
of the 80 or so clusters we searched for cavities, 16 were found
to have them.  If one assumes, for the moment, that all central cluster
galaxies produce bubbles at  a similar rate, then we are 
seeing clusters in an on state only $\sim 20\%$
of the time.  Therefore, the average time elapsed between the production
of bubble pairs could be as large as $5\times 10^8$ yr, at least
in some clusters.  Our search did not discriminate between cooling
and noncooling clusters, and it is likely, although it has not been
proved, that bubbles are produced preferentially in cooling flows.  
If true, our estimate of $\sim 10^8~{\rm yr}$ between outbursts
is probably closer to the truth for most cooling flow clusters.
Then the time averaged  mechanical luminosity discussed below has
been overestimated in most objects by factors
of $2-3$.  On the other hand, the 333 MHz radio map of Hydra~A \citep{lane04}
shows 2' plumes extending several times further from the AGN than the
X-ray cavities (and the 4~GHz radio image).  It also shows an outer
radio lobe 4' north of the AGN that coincides with a feature in
the X-ray image that hints of a distant cavity.  Interpreted
as above, the outer feature would give a very long interval ($\sim10^9$~yr) 
between bubbles, but the plume indicates that such an outer bubble
has been followed (perhaps some time later) by an extended period of
continuous radio activity.  In that case the cavities may just be the
latest in a series, but we are failing to detect most of the remnants.
This would make our estimate of $L_{\rm mech}$ closer to the truth.

\section{Discussion}

\subsection{Trends with Radio Luminosity}

In Figure~\ref{fig-Lr_Lm}, we present two plots showing 
the mechanical luminosity versus the total radio luminosity (\emph{left}) and
the total radio luminosity versus the ratio of the mechanical 
luminosity to the monochromatic, 1.4~GHz radio luminosity (\emph{right}).
In each plot we distinguish between radio-filled and 
ghost cavities, shown with filled symbols and empty symbols
respectively.  The ``error bars'' for each point reflect the range of instantaneous
mechanical luminosity implied by the range in possible ages.
The data are taken or are derived from Tables~\ref{tbl-cluster-radio} and \ref{tbl-bubbles}.  

The left-hand panel of Figure~\ref{fig-Lr_Lm} shows a trend between 
the radio luminosity  and mechanical luminosity, with the sense that 
more luminous radio sources tend toward larger mechanical 
luminosities.   This trend seems to be shared by both the
radio-filled cavities and the ghost cavities, in spite of the 
use of the current central radio power for both the filled cavities
and the ghosts, to which the current central source may be
unrelated. No segregation by FOM is seen.  
The relation between the two luminosities appears 
to be roughly a power law. To quantify this relation, we used a linear 
least-squares fit to the logarithms of the data, with errors in 
mechanical luminosity given by the extreme values for each system. We show in 
Figure~\ref{fig-Lr_Lm} the best-fit line for the entire sample 
(\emph{dashed line}), given by
\begin{equation}
L_{\rm{mech}} = 10^{25\pm 3} \left( L_{\rm{radio}} \right)^{0.44\pm 0.06},
\end{equation}
and for the radio-filled cavities only (\emph{dotted line}), given by
\begin{equation}
L_{\rm{mech}} = 10^{18\pm 4} \left( L_{\rm{radio}} \right)^{0.6\pm 0.1}.
\end{equation}
In both cases, the mechanical luminosity scales as the radio luminosity to 
approximately the one-half power over six decades of radio
power, albeit with large scatter.  

The relative contribution of cosmic scatter and observational uncertainty
is hard to judge without precision radio data at a variety of wavelengths
and without a better understanding of the bubble production timescale.
Nevertheless, the existence of this trend demonstrates quantitatively
that the radio sources are indeed creating
the cavities.  The radio sources are not simply filling preexisting
voids in the intracluster medium (ICM) created by other processes.  Furthermore, the 
synchrotron luminosity and mechanical luminosity do not scale in direct
proportion to each other.  This relationship implies that the 
synchrotron luminosity
cannot be used to infer the mechanical power of a radio jet in
a simple fashion.  

An important and poorly understood aspect
of radio source physics is the degree of coupling
between the mechanical (kinetic) luminosity of radio
sources and their synchrotron luminosity.
This coupling is theoretically tied to the magnetic field 
strength and age of the source 
\citep[see ][]{ddy93,bick97}, neither of which can be
measured reliably from radio data alone.  Radio sources are
inefficient radiators.  The ratio of
mechanical power to radio power is typically assumed to 
range between 10 and 100, almost entirely on theoretical
considerations \citep[]{ddy93,bick97}.  On the other hand, measurements of the
the X-ray cavity sizes and surrounding gas pressures provide
unique estimates of their ages and mechanical luminosities,
independently 
of the radio properties themselves. We evaluate the ratio of mechanical energy
to radio power by plotting the ratio of mechanical power in 
the bubbles to monochromatic, 1.4~GHz synchrotron luminosity, 
assuming $1pV$ of energy per radio lobe, against
the radio luminosity in the right-hand panel of Figure~\ref{fig-Lr_Lm}.  
This ratio ranges from a few to 
a few hundred for the powerful sources, which is broadly consistent
with theoretical estimates \citep[see ][]{ddy93, bick97}.  On the other
hand, Abell~478 has a ratio exceeding a few thousand.
To the extent that X-ray cavities provide a good measure of
the mechanical energy of radio sources, the large variation in this ratio 
indicates that radio luminosity is not necessarily a reliable
probe of the available mechanical energy.

There are several factors that can introduce scatter into
our estimate of the ratio of
radio to kinetic power.  The most important is probably 
intrinsic differences between the radio sources themselves,
a consequence of dramatic changes in radio luminosity with time. 
Certainly, if radio outbursts are to compensate for
radiative losses in cooling flows, then the absence of radio emission
from some systems requires large variations of radio luminosity with
time.  On the other hand,  the $pV$ energy of the bubbles alone 
would tend to underestimate the the mechanical luminosity of
radio sources by factors of several if energy dissipating
shocks are generated, or if the bubbles expand non-adiabiatically (they
leak),  or if the internal energy of the bubbles is
boosted with a relativistic plasma.

\subsection{Heating by Radio-Induced Cavities}\label{sec-heating}

\citet{chur02} noted the conversion of enthalpy of the rising
bubble into other forms of energy in the cluster atmosphere.  Here it is shown that,
for an adiabatic bubble, this energy is dissipated in its wake.  If
the mass in the bubble is negligible compared to the mass of the gas
it displaces, then a bubble rises, because the gas falls in around it
to fill the space it occupied.  This process is driven by the
potential energy released as the surrounding gas moves inward.  
The energy is first converted to gas kinetic energy, then dissipated
in the wake of the rising bubble.  In the notation of \S~\ref{sec-cavage}, 
the potential energy released when the bubble rises a
small distance, $\delta R$, is 
\begin{equation}
\delta W = \rho V g \, \delta R = - V {dp\over dR} \, \delta R,
\end{equation}
where $\rho$ is the gas density, and we have used the equation of
hydrostatic equilibrium to replace $\rho g = -dp/dR$, where $p$ is the
gas pressure.  This gives a differential equation for the energy
dissipated in the bubble wake 
\begin{equation}
{dW\over dR} = - V {dp\over dR}.
\end{equation}
If the bubble is adiabatic, with ratio of specific heats $\gamma$,
then $p V^\gamma =$ constant and this equation can be integrated to
give the energy dissipated as the bubble rises over a large distance,
from $R_0$ to $R_1$,
\begin{equation}
\Delta W ={\gamma\over \gamma - 1} ( p_0 V_0 - p_1 V_1) = H_0 - H_1.
\end{equation}
Here, subscripts 0 and 1 label quantities at the corresponding radii,
and the enthalpy of the bubble is $H = \gamma p V /(\gamma - 1)$.
Note that the bubble is assumed to be
small compared to $R$ (otherwise there can be a significant change in
the density of the gas as it falls in around the bubble).  This would
rarely be significant in a cluster, but when it is, then some of
the potential energy 
goes into readjustment of the atmosphere as the bubble moves.

For a relativistic gas, $\gamma = 4/3$, so that the enthalpy is $4pV$.
The region where this is dissipated by an adiabatic bubble is
determined by the pressure distribution of the atmosphere.  For
clusters such as Hydra~A and 
Perseus, roughly half of this energy would be dissipated inside the cooling
radius.  It is likely that the bubbles are not entirely
adiabatic.  On the basis of our numbers, radio losses are generally
negligible, but pieces may be broken away from bubbles, and the
relativistic particles may leak.  Such effects will generally lead to
a greater proportion of the bubble energy being deposited within the
cooling radius. 

It is important to note that our estimate of the mechanical
luminosity relies critically on the assumption that the bubbles are
close to local pressure equilibrium.  This is at least approximately
true for the Hydra~A cluster \citep{nuls02}.  However, according to
the standard view of radio sources, bubbles may have been
significantly overpressured while being formed \citep[e.g.\
][]{hein98}.  In that case, the expanding bubble drives a shock, and the
energy deposited by the expansion can be substantially larger than $pV$.

There may be additional heat input from the AGNs associated with radio
outbursts.  This could take the form of spherical shocks (driven by
poorly collimated outflows), direct injection of relativistic
particles, inverse Compton heating \citep{ciot01}, or other processes.
Very substantial additional heat inputs would drive convection,
leading to an isentropic core and mixing out of abundance gradients
\citep{brug02a}, but this is not a very strong constraint.  If such energy
injection is significantly more than the bubble energy input, then it
is inappropriate to associate it directly with the bubbles, but the
mean heating power may be correlated with bubble mechanical power.

Finally, it should be noted that even for adiabatic bubbles, the free
energy of a bubble decreases with time, and bubbles may even break up
quickly, so that they disappear as X-ray cavities.  This means that
the instantaneous estimate of bubble mechanical power that we have used
varies with time, and may vary dramatically.  A much better controlled
sample is needed to investigate such issues.

\phn

\subsection{Can Cavity Production Quench Cooling Flows?}\label{sec-quenching}

\begin{figure}
\plotone{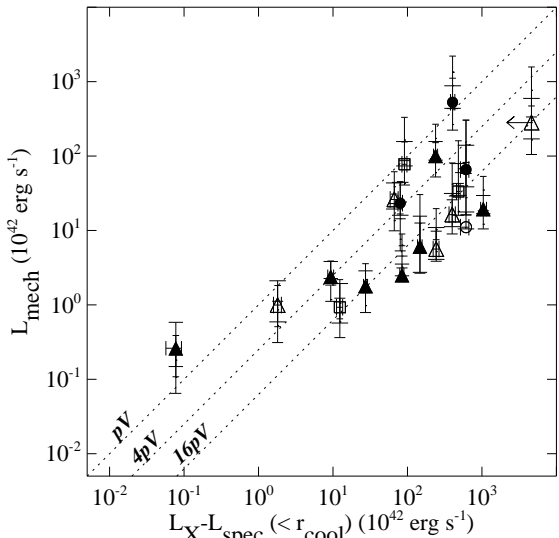}
\caption{Mechanical luminosity vs.\ total luminosity minus the 
spectroscopic estimate of the cooling luminosity.  Lines denoting 
$L_{\rm{X}} - L_{\rm{spec}} = L_{\rm{mech}}$ are 
shown for the assumption of $pV,$ $4pV,$ and $16pV$ 
energy in the bubbles.  Symbols and error bars as in
Figure~\ref{fig-Lr_Lm}; the arrow denotes an upper limit.   \label{fig-Lx_Lm}}
\end{figure}

We now turn to the question of whether radio sources deposit enough
energy into the ICM to quench cooling.  
We use $L_{\rm{X}}$, the total luminosity of the X-ray--emitting
gas from within the cooling radius, as an estimate of the 
classical, or morphological,
cooling luminosity in the absence of heating 
and $L_{\rm{spec}}$, the spectral 
estimate of the cooling luminosity within the cooling radius, as the luminosity
of the gas cooling to low temperatures.  The cooling luminosity, 
$L_{\rm{X}} - L_{\rm{spec}}$, must be offset by heating 
in order to prevent the gas from cooling to low temperatures.
We note that this quantity ignores non--X-ray cooling, such as ultraviolet 
and optical emission, predicted to result from cooling by thermal 
conduction inside magnetic flux loops \citep{sok04}, or along reconnected 
magnetic field lines between cold clouds and the ICM \citep{sok03}. 
Any such emission would lower the 
cooling luminosity which must be balanced by heating.
Figure~\ref{fig-Lx_Lm} shows the mechanical
luminosity plotted against $L_{\rm{X}} - L_{\rm{spec}}$ for our sample.
The diagonal lines represent equality between cooling and
heating, assuming energy inputs of $pV,$ $4 pV,$ and $16 pV$ per cavity.  
The data are derived from Tables~\ref{tbl-cluster-xray} and \ref{tbl-bubbles}.  
For RBS~797, an upper limit is shown.  The cooling luminosity for RBS~797 is 
poorly constrained by the spectrum, which consists of only $\sim 9000$ counts 
after cleanin.  RBS~797, while very luminous, is the most distant cluster in 
our sample and has the shortest exposure time (see Section~\ref{sec-sample}).

Figure~\ref{fig-Lx_Lm} shows several objects, such as Hydra~A, Cygnus~A, and M84,
whose cavities can contain enough energy to 
balance radiative losses, at least temporarily,  
with nearly $1pV$ of heat input per cavity.
The remaining objects, which require 
between a few and $\sim 20 pV$ per cavity to balance cooling, would
do so with varying degrees of difficulty.  As discussed above, $\sim 2
pV$ would be deposited within the cooling radius by an adiabatic
bubble containing relativistic plasma.  Up to $4 pV$ is available if
the cavities are relativistic and nonadiabatic, and there may be
further energy input if they are overpressured or produce a
shock when they are formed.  Therefore, the objects that require $\sim
4 pV$ or less may reasonably be supplied with enough energy in the cavities
to balance cooling, depending on the detailed dynamics
(the heat also needs to be distributed inside the cooling radius to
match the distribution of radiative losses).
Provided that the true radio cycling timescale ranges between
$t_{\rm{c_{s}}}$ and $t_{\rm r}$, the cavities in one quarter to
one-half of the objects in our sample contain enough
energy to offset radiation losses.
This would be true of the cavities in the remaining objects only if
they are significantly nonadiabatic, as
outlined in \S~\ref{sec-quenching}.  Bear in mind that our conclusions depend on
the adopted cooling radius (see \S 3.2), measurement uncertainties
in the cavity sizes, and the cavity production timescale.
Nevertheless, we can safely conclude that cooling can plausibly be balanced by
bubble heating in some, but not all, systems.

It is unnecessary to balance the entire 
luminosity, $L_{\rm{X}} - L_{\rm{spec}},$ 
by bubble heating alone if there are other forms of heating present.  
A possible source of heating is thermal conduction, which, as demonstrated by \citet{voig04}, 
could supply a significant amount of heat.  Using a sample similar
to our own, Voigt \& Fabian found that thermal conduction can
reduce the cooling luminosity by factors of $\sim 2-3$ in
some objects. Although they are difficult to find in X-ray images, 
shocks associated with the expanding cavities
can deposit additional energy into the ICM.  Deep {\it Chandra}
images of a growing number of objects, including 
Cygnus~A \citep{wils03}, NGC 4636 \citep{jones02}, M87 \citep{form04}, 
and Perseus \citep{fabi03a}, show surface brightness discontinuities
that may be associated with weak shocks.  
In Cygnus~A and M87, the shocks imply that the radio source may provide 
several times the upper limit of the luminosity seen in the bubbles, under the 
assumption of $4pV$ of energy per bubble \citep{wils03,form04}. 
It may therefore be a combination of heating mechanisms that 
leads to quenched cooling, as suggested by several 
authors \citep{brig02,brig03,kim03,rusz02}.

It is important to note that that our sample is biased toward
systems with visible evidence for X-ray cavities, and does not
represent clusters as a whole.  Many clusters, including some with
large cooling flows, do not contain cavities \citep[e.g.\ Abell 1068,][]{wise03,mcna04}.
These objects may have very different reheating histories than
the objects discussed here.  In this sense, the objects presented
here represent the best-case examples for reheating the ICM
by energetic bubbles.  Our analysis does not imply that all
cooling flows can be quenched in this fashion.

\subsection{Trends between X-ray and Mechanical Luminosities}\label{sec-Lx-trends}

Figure~\ref{fig-Lx_Lm} shows a trend between the X-ray
luminosity and bubble mechanical luminosity, with
the sense that systems with larger X-ray luminosities also have larger
mechanical luminosities.  This trend 
extends over a dynamic range of $\sim 1000$ in both X-ray
and mechanical luminosity.  Just such a trend would be expected were the 
cooling and heating of the ICM coupled in some fashion.
Several studies \citep[e.g.][]{rosn89,binn95,davi01,quil01,chur02}
have proposed that cooling is balanced by heating 
in a self-regulated feedback loop.  The feedback loop is driven by
episodic radio activity fueled by cooling and accretion 
onto a central black hole.  The accretion
energy is then returned to the ICM through an AGN outburst, including
the action of the radio cavities, which temporarily arrests 
cooling.  At later times, the center of the system settles down and the cooling 
flow is reestablished.  During the cooling cycle,  molecular gas \citep{edge01} accumulates
and star formation ensues \citep{jfn,mo89}, 
albeit at substantially lower levels
than expected in steady-cooling models \citep{fabi94}.
Even if the radio bubbles are not the main source of heat, 
Figure~\ref{fig-Lx_Lm} suggests that AGN 
feedback is intimately involved in the process
that prevents a large cooling flow from forming.

The apparent correlation in Figure~\ref{fig-Lx_Lm}
should be treated with caution. 
As noted earlier, our sample was selected from
clusters in the \textit{Chandra} archive with fairly obvious
cavities in their cores, and neglects those without obvious cavities.
Other clusters are known to have substantial cooling luminosities
commensurate with the observed 
levels of cold gas and star formation, yet contain no cavities
and have low radio power.  A prime example is A1068 
\citep{wise03,mcna04}.
Similar objects would appear
in the lower right of this diagram, tending to
weaken the correlation.  On the other hand, it is unlikely
that we would have missed objects with powerful
cavities, which would lie in the upper part 
of this diagram.  Therefore, the distribution of points may
represent an upper envelope in mechanical luminosity as a function
of  X-ray luminosity.  Such a distribution would be consistent with
the feedback hypothesis, if objects like A1068 
are in an extended cooling phase in which the central 
galaxies have experienced substantial levels of accretion in the past
100~Myr or so, when the radio source has not had a chance to create 
cavities capable of reducing or quenching cooling.

We have investigated the degree to which other systematic effects
may lead to an unphysical luminosity-luminosity correlation.  
For example, \citet{elvi78} pointed out that a sample of 
objects with a small range of fluxes and a large range of distances 
will show a correlation in a luminosity-versus-luminosity plot, 
even if there is no intrinsic correlation in the sample.  Our 
sample, however, has a large range of radio fluxes 
(from $\sim 10$~mJy to $\sim 10^{6}$~mJy).  
The cavities in our sample also cover a large 
range of projected 
angular sizes (from $\sim 3$" to $\sim 35$").  We believe, then,
that these potential effects are unlikely to account entirely 
for the trends seen in Figures~\ref{fig-Lr_Lm} and \ref{fig-Lx_Lm}.

Selection bias may also contribute to the correlations.  
Small cavities are easily overlooked in distant objects, 
since cavities of a given linear size become
more difficult to detect as their angular sizes decrease with
increasing distance.
Conversely, in very nearby objects, such as M87, we may miss 
larger bubbles that lie outside the detector.  Furthermore, 
other considerations, such as the bubble position, affect the 
detectability \citep{enss02}.  The consequences of these and
other effects on our selection function will be addressed
in the future using a larger and better-defined sample of clusters, 
including a more sophisticated approach to placing limits on cavities
that may exist in clusters but were missed by the observations.

\subsection{Summary}

We have presented an analysis of 18 systems taken from 
the \textit{Chandra} archive having clear evidence for cavities
in their X-ray emission.  We find that the energy associated 
with the cavities is sufficient to 
substantially reduce or quench cooling in nearly half of the
objects in our sample.  However, this mechanism alone probably does not 
provide a general solution to the cooling problem, unless X-ray
cavities probe only a small fraction of the total kinetic luminosity
of radio sources. In addition, we have discovered a trend between the
cooling X-ray luminosity and the mechanical energy of the cavities,
with the sense that more luminous systems produce larger and
more energetic cavities. The  trend, or envelope, may
have been established by a self-regulated cooling and feedback mechanism
acting in many systems. The existence of such a mechanism
in relatively nearby clusters, where the detailed physics can
be examined, may provide significant insight on the process of galaxy formation
that prevails at large redshifts \citep[e.g.\ ][]{voit03}.
A similar mechanism may regulate the growth of galaxy halos 
during the dissipative
stages of their development \citep{dubi94}
and may be an agent responsible
for the detailed correlation between black hole mass and velocity
dispersion of spheroids \citep{fabi02b}.
We have measured for the first time the distribution 
of the ratio of kinetic luminosity
to monochromatic radio luminosity for a sample of radio sources.
The ratio varies widely, with most objects ranging between few and
a few hundred, assuming $1pV$ of energy per cavity.  X-ray cavities
provide a unique probe of the mechanical power of radio jets, 
independently of the radio properties themselves. 

Our future plans include expanding the sample size
and acquiring better and more uniform radio data.  
In addition, we plan to extend our 
understanding of the detectability function of bubbles, using simulations of
images with a wide range of exposure and
signal-to-noise ratios.

\phn

We acknowledge helpful discussions with Mangala Sharma, Liz Blanton,
Frazer Owen, Hui Li, Phil Kronberg, and David De Young, and we thank Carlo 
Nipoti for pointing out an error in the Centaurus radio flux in an earlier version. 
This work was supported by Long Term Space
Astrophysics Grant NAG5-11025, \textit{Chandra} Archive Grant AR2-3007X, and a grant
from the Department of Energy through the Los Alamos National Laboratory.


\begin{thebibliography}{}
\bibitem[Basson \& Alexander(2003)]{bass03} Basson, J. F., \& Alexander, P.  2003, \mnras, 339, 353
\bibitem[Becker et al.(1991)]{beck91} Becker, R. H., White, R. L., \& Edwards, A. L.  1991, \apjs, 75, 1
\bibitem[Bender, Saglia, \& Gerhard(1994)]{bend94} Bender, R., Saglia, R. P., \& Gerhard, O. E. 1994, \mnras, 269, 785
\bibitem[Bertschinger \& Meiksin(1986)]{bert86} Bertschinger, E., \& Meiksin, A.  1986, \apj, 306, L1
\bibitem[Bicknell et al.(1997)]{bick97} Bicknell, G. V., Dopita, M. A., \& O'Dea, C. P. 1997, \apj, 485, 112
\bibitem[Binney(2004)]{binn03} Binney, J.  2004, in The Riddle of Cooling Flows in Galaxies and Clusters of Galaxies, ed. T. H. Reiprich, J. C. Kempner, \& N. Soker (Charlottesville: Univ. Virginia), http://www.astro.virginia.edu/coolflow
\bibitem[Binney \& Tabor(1995)]{binn95} Binney, J., \& Tabor, G.  1995, \mnras, 276, 663
\bibitem[Binney \& Tremaine(1987)]{binn87} Binney, J., \& Tremaine, S.  1987, Galactic Dynamics (Princeton: Princeton Univ. Press)
\bibitem[Blakeslee \& Tonry(1992)]{blak92} Blakeslee, J. P., \& Tonry, J. L.  1992, \aj, 103, 1457
\bibitem[Blanton et al.(2003)]{blan03} Blanton, E. L., Sarazin, C. L., \& McNamara, B. R.  2003, \apj, 585, 227
\bibitem[Blanton et al.(2001)]{blan01} Blanton, E. L., Sarazin, C. L., McNamara, B. R., \& Wise, M. W.  2001, \apj, 558, L15
\bibitem[B\"{o}hringer \& Hensler(1989)]{boeh89} B\"{o}hringer, H., Hensler, G.  1989, \aap, 215, 147
\bibitem[B\"{o}hringer et al.(2002)]{boeh02} B\"{o}hringer, H., Matsushita, K., Churazov, E., Ikebe, Y., \& Chen, Y.  2002, \aap, 382, 804
\bibitem[B\"{o}hringer \& Morfill(1988)]{boeh88} B\"{o}hringer, H., \& Morfill, G. E.  1988, \apj, 330, 609
\bibitem[B\"{o}hringer et al.(1993)]{boeh93} B\"{o}hringer, H., Voges, W., Fabian, A. C., Edge, A. C., Neumann, D. M. 1993, \mnras, 264, 25
\bibitem[Bregman \& David(1988)]{bregdav} Bregman, J. N., \& David, L. P. 1988, ApJ, 326, 639
\bibitem[Brighenti \& Matthews(2002)]{brig02} Brighenti, F., \& Matthews, W. G.  2002, \apj, 573, 542
\bibitem[Brighenti \& Matthews(2003)]{brig03} Brighenti, F., \& Matthews, W. G.  2003, \apj, 587, 580
\bibitem[Br\"{u}ggen(2002)]{brug02a} Br\"{u}ggen, M.  2002, \apj, 571, L13
\bibitem[Br\"{u}ggen(2003)]{brug03}  Br\"{u}ggen, M.  2003, \apj, 592, 839
\bibitem[Br\"{u}ggen \& Kaiser(2001)]{brug01} Br\"{u}ggen, M., \& Kaiser, C. R.  2001, \mnras, 325, 676
\bibitem[Br\"{u}ggen et al.(2002)]{brug02b} Br\"{u}ggen, M., Kaiser, C. R., Churazov, E., \& En\ss lin, T. A.  2002, \mnras, 331, 545
\bibitem[Burbidge \& Crowne(1979)]{burb79} Burbidge, G., \& Crowne, A. H.  1979, \apjs, 40, 583
\bibitem[Carilli et al.(1994)]{carilli} Carilli, C. L., Perley, R. A., \& Harris, D. E. 1994, MNRAS, 270, 173
\bibitem[Carollo et al.(1993)]{caro93} Carollo, C. M., Danzinger, I. J., \& Buson, L.  1993, \mnras, 265, 553
\bibitem[Carter et al.(1985)]{cart85} Carter, D., Inglis, I., Ellis, R. S., Efstathiou, G., \& Godwin, J. G.  1985, \mnras, 212, 471
\bibitem[Churazov et al.(2001)]{chur01} Churazov, E., Br\"{u}ggen, M., Kaiser, C. R., B\"{o}hringer, H., \& Forman, W. R.  2001, \apj, 554, 261
\bibitem[Churazov et al.(2002)]{chur02} Churazov, E., Sunyaev, R., Forman, W., \& B\"{o}hringer, H., 2002, \mnras, 332, 729
\bibitem[Ciotti \& Ostriker(2001)]{ciot01} Ciotti, L., \& Ostriker, J. P. 2001, \apj, 551, 131
\bibitem[Condon et al.(1998)]{cond98} Condon, J. J., Cotton, W. D., Greisen, E. W., Yin, Q. F., Perley, R. A., Taylor, G. B., \& Broderick, J. J.  1998, \aj, 115, 1693
\bibitem[David et al.(2001)]{davi01} David, L. P., Nulsen, P. E. J., McNamara, B. R., Forman, W., Jones, C., Ponman, T., Robertson, B., \& Wise, M. W.  2001, \apj, 557, 546
\bibitem[De Young(1993)]{ddy93} De Young, D. S. 1993, \apj, 405, L13
\bibitem[De Young(2003)]{ddy03} De Young, D. S. 2003, \mnras, 343, 219
\bibitem[Dolag et al.(2004)]{dola04} Dolag, K., Jubelgas, M., Springel, V., Borgani, S., \& Rasia, E.  2004, \apjl, submitted (astro-ph/0401470)
\bibitem[Dubinski(1994)]{dubi94} Dubinski, J.  1994, \apj, 431, 617
\bibitem[Edge(2001)]{edge01} Edge, A. C.  2001, \mnras, 328, 762
\bibitem[Elvis et al.(1978)]{elvi78} Elvis, M., Maccacaro, T., Wilson, A. S., Wark, M. J., Penston, M. V., Fosbury, R. A., \& Perola, G. C., 1978, \mnras, 183, 129
\bibitem[En\ss lin \& Heinz(2002)]{enss02} En\ss lin, T. A., \& Heinz, S.  2002, \aap, 384, L27
\bibitem[Ettori et al.(2002)]{etto02} Ettori, S., Fabian, A. C., Allen, S. W., \& Johnstone, R. M.  2002, \mnras, 331, 635
\bibitem[Fabian(1994)]{fabi94} Fabian, A. C.  1994, \araa, 32, 277
\bibitem[Fabian et al.(2002a)]{fabi02a} Fabian, A. C., Celotti, A., Blundell, K. M., Kassim, N. E., \& Perley, R. A.  2002a, \mnras, 331, 369
\bibitem[Fabian et al.(2001)]{fabi01} Fabian, A. C., Mushotzky, R. F., Nulsen, P. E. J., \& Peterson, J. R.  2001, \mnras, 321, L20
\bibitem[Fabian et al.(2003a)]{fabi03a} Fabian, A. C., Sanders, J. S., Allen , S. W., Crawford, C. S., Iwasawa, K., Johnstone, R. M., Schmidt, R. W., \& Taylor, G. B.  2003a, MMRAS, 344, L43 
\bibitem[Fabian et al.(2003b)]{fabi03b} Fabian, A. C., Sanders, J. S., Crawford, C. S., Conselice, C. J., Gallagher III, J. S., \& Wyse, R. F. G.  2003b, MNRAS, 344, L48
\bibitem[Fabian et al.(2002b)]{fabi02b} Fabian, A. C., Wilman, R. J., \& Crawford, C. S.  2002, \mnras, 329, L18
\bibitem[Fabian et al.(2000)]{fabi00} Fabian, A. C., Sanders, J. S., Ettore, S., Taylor, G. B., Allen, S. W., Crawford, C. S., Iwasawa, K., Johnstone, R. M., \& Ogle, P. M.  2000, \mnras, 318, L65
\bibitem[Finoguenov \& Jones(2001)]{fino01} Finoguenov, A., \& Jones, C.  2001, \apj, 547, L107
\bibitem[Fisher et al.(1995)]{fish95} Fisher, D., Illingworth, G., \& Franx, M.  1995, \aj, 438, 539
\bibitem[Forman et al.(2004)]{form04} Forman, W., Nulsen, P. E. J., Heinz, S., Owen, F., Eilek, J., Vikhlinin, A., Markevitch, M., Kraft, R., Churazov, E., \& Jones, C.  2004, ApJ, submitted (astro-ph/0312576)
\bibitem[Fujita et al.(2002)]{fuji02} Fujita, Y., Sarazin, C. L., Kempner, J. C., Rudnick, L., Slee, O. B., Roy, A. L., Andernach, H., \& Ehle, M.  2002, \apj, 575, 764
\bibitem[Gull \& Northover(1973)]{gull73} Gull, S. F., \& Northover, K. J. E.  1973, Nature, 244, 80
\bibitem[Heckman et al.(1985)]{heck85} Heckman, T. M., Illingworth, G. D., Miley, G. K., \& van Breugel, W. J. M.  1985, \aj, 299, 41
\bibitem[Heinz et al.(2002)]{hein02} Heinz, S., Choi, Y., Reynolds, C. S., \& Begelman, M. C.  2002, \apj, 569, L79
\bibitem[Heinz et al.(1998)]{hein98} Heinz, S., Reynolds, C. S., \& Begelman, M. C.  1998, \apj, 501, 126
\bibitem[Johnstone et al.(2002)]{john02} Johnstone, R. M., Allen, S. W., Fabian, A. C., \& Sanders, J. S.  2002, \mnras, 336, 299
\bibitem[Johnstone et al.(1987)]{jfn} Johnstone, R. M, Fabian, A. C., \& Nulsen, P. E. J. 1987, MNRAS, 224, 75
\bibitem[Jones et al.(2002)]{jones02} Jones, C., Forman, W., Vikhlinin, A., Markevitch, M., David, L., Warmflash, A., Murray, S., \& Nulsen, P. E. J. 2002, ApJ, 567, L115
\bibitem[Kaastra et al.(2004)]{kaast04} Kaastra, J.S., Tamura, T., Peterson, J. R., Bleeker, J. A. M., Ferrigno, C., Kahn, S. M., Paerels, F. B. S., Piffaretti, R., Branduardi-Raymont, G., \& B\"ohringer, H. 2004, \aap, 431, 415
\bibitem[Kaiser \& Binney(2003)]{kais03} Kaiser, C. R., \& Binney, J.  2003, \mnras, 338, 837
\bibitem[Kim \& Narayan(2003)]{kim03} Kim, W., \& Narayan, R.  2003, \apj, 596, 889
\bibitem[K\"{u}hr et al.(1981)]{kueh81} K\"{u}hr, H., Witzel, A., Pauling-Toth, I. I. K., \& Nauber, U.  1981, \aaps, 45, 367
\bibitem[Lane et al.(2004)]{lane04} Lane, W. M., Clarke, T. E., Taylor, G. B., Perley, R. A., \& Kassim, N. E.  2004, \aj, 127, 48
\bibitem[Loewenstein et al.(1991)]{loew91} Loewenstein, M., Zweibel, E. G., \& Begelman, M. C.  1991, \apj, 377, 392
\bibitem[Makishima et al.(2001)]{maki01} Makishima, K., Ezawa, H., Fukuzawa, Y. Honda, H., Ikebe, Y., Kamae, T., Kikuchi, K., Matsushita, K., Nakazawa, K., Ohashi, T., Takahashi, T., Tamura, T., \& Xu, H.  2001, \pasj, 53, 401
\bibitem[Mathews et al.(2003)]{math03} Mathews, W. G., Brighenti, F., Buote, D. A., \& Lewis, A. D.  2003, \apj, 596, 159
\bibitem[Mazzotta et al.(2003)]{mazz03} Mazzotta, P., Edge, A., \& Markevitch, M.  2003, \apj, 596, 190
\bibitem[Mazzotta et al.(2002)]{mazz02} Mazzotta, P., Kaastra, J. S., Paerels, F. B., Ferrigno, C., Colafrancesco, S., Mewe, R., \& Forman, W. R.  2002, \apj, 567, L37
\bibitem[McNamara \& O'Connell(1989)]{mo89} McNamara, B. R., \& O'Connell, R.W. 1989, AJ, 98, 2018
\bibitem[McNamara et al.(2004)]{mcna04} McNamara, B. R., Wise, M. W., \& Murray, S. S.  2004, \apj, 601, 173
\bibitem[McNamara et al.(2000)]{mcna00} McNamara, B. R., Wise, M. W., Nulsen, P. E. J., David, L. P., Sarazin, C. L., Bautz, M., Markevitch, M., Vikhlinin, A, Forman, W. R., Jones, C., \& Harris, D. E.  2000, \apj, 534, L135
\bibitem[McNamara et al.(2001)]{mcna01} McNamara, B. R., Wise, M. W., Nulsen, P. E. J., David, L. P., Carilli, C. L., Sarazin, C. L., O'Dea, C. P., Houck, J., Donahue, M., Baum, S., Voit, M., O'Connell, R. W., \& Koekemoer, A.  2001, \apj, 562, L149
\bibitem[Molendi \& Pizzolato(2001)]{mole01} Molendi, S., \& Pizzolato, F.  2001, \apj, 560, 194
\bibitem[Nulsen et al.(2002)]{nuls02} Nulsen, P. E. J., David, L. P., McNamara, B. R., Jones, C., Forman, W. R., \& Wise, M. W.  2002, \apj, 568, 163
\bibitem[Omma et al.(2004)]{omma04} Omma, H., Binney, J., Bryan, G., \& Slyz, A.  2004, \mnras, 348, 1105
\bibitem[Pedlar et al.(1990)]{pedl90} Pedlar, A., Ghataure, H. S., Davies, R. D., Harrison, B. A., Perley, R., Crane, P. C., \& Unger, S. W.  1990, \mnras, 246, 477
\bibitem[Peterson et al.(2003)]{pete03} Peterson, J. R., Kahn, S. M., Paerels, F. B., Kaastra, J. S., Tamura, T., Bleeker, J. A. M., Ferrigno, C., \& Jernigan, J. G.  2003, \apj, 590, 207
\bibitem[Peterson et al.(2001)]{pete01} Peterson, J. R., Paerels, F. B., Kaastra, J. S., Arnaud, M., Reiprich, T. H., Fabian, A. C., Mushotzky, R. F., Jernigan, J. G. 2001, \& Sakelliou, I. \aap, 365, L104
\bibitem[Quilis et al.(2001)]{quil01} Quilis, V., Bower, R. G., \& Balogh, M. L.  2001, \mnras, 328, 1091
\bibitem[Reynolds, Heinz, \& Begelman(2002)]{reyn02} Reynolds, C. S., Heinz, S., \& Begelman, M. C.  2002, \mnras, 332, 271
\bibitem[Robinson et al.(2004)]{robi04} Robinson, K., Dursi, L. J., Ricker, P. M., Rosner, R., Calder, A. C., Zingale, M., Truran, J. W., Linde, T., Caceres, A., Fryxell, B., Olson, K., Riley, K., Siegel, A., \& Vladimirova, N.  2004, \apj, 601, 621
\bibitem[Rosner \& Tucker(1989)]{rosn89} Rosner, R., \& Tucker, W. H.  1989, \apj, 338, 761
\bibitem[Ruszkowski \& Begelman(2002)]{rusz02} Ruszkowski, M., \& Begelman, M. C.  2002, \apj, 581, 223
\bibitem[Sanders \& Fabian(2002)]{sand02} Sanders, J. S., \& Fabian, A. C.  2002, \mnras, 331, 273
\bibitem[Sarazin et al.(1995a)]{sara95a} Sarazin, C. L., Baum, S. A., \& O'Dea, C.  1995, \apj, 451, 125
\bibitem[Sarazin et al.(1995b)]{sara95b} Sarazin, C. L., Burns, J. O., Roettiger, K., \& McNamara, B. R.  1995, \apj, 447, 559
\bibitem[Schindler et al.(2001)]{schi01} Schindler, S., Castillo-Morales, A., De Filippis, E., Schwope, A., \& Wambsganss, J.  2001, \aap, 376, L27
\bibitem[Schmidt et al.(2002)]{schm02} Schmidt, R. W., Fabian, A. C., \& Sanders, J. S.  2002, \mnras, 337, 71
\bibitem[Slee(1995)]{slee95} Slee, O. B.  1995, AuJPh, 48, 143
\bibitem[Slee et al.(2001)]{slee01} Slee, O. B., Roy, A. L., Murgia, M., Andernach, H., \& Ehle, M.  2001, \aj, 122, 1172 
\bibitem[Slee \& Siegman(1988)]{slee88} Slee, O. B., \& Siegman, B. C.  1988, \mnras, 235, 1313
\bibitem[Smith et al.(2002)]{smit02} Smith, D. A., Wilson, A. S., \& Arnaud, K. A.  2002, \apj, 565, 195
\bibitem[Smith et al.(1990)]{smit90} Smith, E. P., Heckman, T. M., \& Illingworth, G. D.  1990, \aj, 356, 399
\bibitem[Soker(2003)]{soke03} Soker, N.  2003, \mnras, 342, 463
\bibitem[Soker(2004)]{sok04} Soker, N. 2004, MNRAS, submitted (astro-ph/0311014)
\bibitem[Soker et al.(2004)]{sok03} Soker, N., Blanton, E. L., \& Sarazin, C. L.  2003, in The Riddle of Cooling Flows in Galaxies and Clusters of Galaxies, ed. T. H. Reiprich, J. C. Kempner, \& N. Soker (Charlottesville: Univ. Virginia), http://www.astro.virginia.edu/coolflow
\bibitem[Soker et al.(2001)]{sok01} Soker, N., White, R. E, III, David, L. P., \& McNamara, B. R.  2001, ApJ, 549, 832
\bibitem[Spinrad et al.(1985)]{spin85} Spinrad, H., Djorgovski, S., Marr, J., \& Aguilar, L.  1985, \pasp, 97, 932
\bibitem[Sun et al.(2003)]{sun03} Sun, M., Jones, C., Murray, S. S., Allen, S. W., Fabian, A. C., \& Edge, A. C.  2003, \apj, 587, 619
\bibitem[Tabor \& Binney(1993)]{tabo93} Tabor, G., \& Binney, J.  1993, \mnras, 263, 323
\bibitem[Tamura et al.(2001)]{tamu01} Tamura, T., Kaastra, J. S., Peterson, J. R., Paerels, F. B. S., Mittaz, J. P. D., Trudolyubov, S. P., Stewart, G., Fabian, A. C., Mushotzky, R. F., Lumb, D. H., \& Ikebe, Y.  2001, \aap, 365, L87
\bibitem[Taylor et al.(1994)]{tayl94} Taylor, G. B., Barton, E. J., \& Ge, J.  1994, \aj, 107, 1942
\bibitem[Tonry(1985)]{tonr85} Tonry, J. L.  1985, \aj, 90, 2431
\bibitem[Tucker \& David(1997)]{tuck97} Tucker, W. H., \& David, L. P.  1997, \apj, 484, 602
\bibitem[Tucker \& Rosner(1983)]{tuck83} Tucker, W. H., \& Rosner, R.  1983, \apj, 267, 547
\bibitem[Voigt \& Fabian(2004)]{voig04} Voigt, L. M., \& Fabian, A. C.  2004, \mnras, 347, 1130
\bibitem[Voigt et al.(2002)]{voig02} Voigt, L. M., Schmidt, R. W., Fabian, A. C., Allen, S. W., \& Johnstone, R. M.  2002, \mnras, 335, L7
\bibitem[Voit \& Ponman(2003)]{voit03} Voit, G. M., \& Ponman, T. J.  2003, \apj, 594, L75
\bibitem[von Hoerner(1974)]{vonh74} von Hoerner, S.  1974, in Galactic and Extragalactic Radio Astronomy, ed. G. L. Vershuur \& K. I. Kellermann (Berlin: Springer), 353
\bibitem[Vrtilek et al.(2002)]{vrti02} Vrtilek, J. M, Grego, L., David, L. P., Pomnman, T. J., Forman, W., Jones, C., \& Harris, D. E.  2002, APS Meeting, B17.107, http://www.aps.org/meet/APR02/baps/abs/S200107.html
\bibitem[Wilson et al.(2003)]{wils03} Wilson, A. S., Young, A. J., \& Smith, D. A.  2003, in ASP Conf. Ser. 290, Active Galactic Nuclei: from Central Engine to Host Galaxy, ed. S. Collin, F. Combes, \& I. Shlosman (San Fransico: ASP), 141
\bibitem[Wise et al.(2004)]{wise03} Wise, M. W., McNamara, B. R., \& Murray, S. S.  2004, \apj, 601, 184
\bibitem[Wright et al.(1994)]{wrig94} Wright, A. E., Griffith, M. R., Burke, B. F., \& Ekers, R. D.  1994, \apjs, 91, 111
\bibitem[Wright et al.(1996)]{wrig96} Wright, A. E., Griffith, M. R., Hunt, A. J., Troup, E., Burke, B. F., \& Ekers, R. D.  1996, \apjs, 103, 145
\bibitem[Wright \& Otrupcek(1990)]{wrig90} Wright, A. E., \& Otrupcek, R.  1990, Parkes Radio Sources Catalogue (Epping: ATNF)
\bibitem[Young et al.(2002)]{youn02} Young, A. J., Wilson, A. S., \& Mundell, C. G.  2002, \apj, 579, 560
\bibitem[Zakamska \& Narayan(2003)]{zaka03} Zakamska, N. L., \& Narayan, R.  2003, \apj, 582, 162
\end{thebibliography}
\end{document}